\documentclass[12pt]{iopart}

\usepackage{graphicx}
\usepackage{color}
\usepackage{iopams}

\begin{document}

\title[Advanced approach to the analysis of the W L$_3$-edge EXAFS of tungsten]{Advanced approach to the local structure reconstruction and theory validation on the example of the W L$_3$-edge extended X-ray absorption fine structure of tungsten}

\author{Inga Jonane, Andris Anspoks, Alexei Kuzmin}

\address{Institute of Solid State Physics, University of Latvia, Kengaraga street 8, LV-1063 Riga, Latvia}

\ead{a.kuzmin@cfi.lu.lv}

\begin{abstract}
Atomistic simulations of the experimental W L$_3$-edge extended X-ray absorption fine structure (EXAFS) of bcc tungsten  at $T=300$~K were performed using classical molecular dynamics (MD) and  reverse Monte Carlo (RMC) methods.
The MD-EXAFS method based on the results of MD simulations allowed us to access the structural information, encoded in EXAFS, beyond  the first  coordination shell and to validate the accuracy of two interaction potential models -- the embedded atom model potential and the second nearest-neighbor modified embedded atom method potential. 
The RMC-EXAFS method was used for more elaborate analysis of the EXAFS data giving access to thermal disorder effects. The results of both methods suggest that 
the correlation in atomic motion in bcc tungsten becomes negligible above 8~\AA. This fact allowed us to use the EXAFS data to determine not only mean-square relative displacements of
atomic W--W pair motion but also mean-square displacements of individual tungsten atoms, 
which are usually accessible from diffraction data only.
\end{abstract}

\vspace{2pc}


\submitto{Modelling Simul. Mater. Sci. Eng. 26 (2018) 025004 (11 pp), doi:10.1088/1361-651X/aa9bab}

\maketitle


\section{Introduction}
\label{intro}

Tungsten and its alloys are important materials for plasma-facing components in fusion reactors, which are expected to withstand sever damage of their microstructure when exposed to high-energy irradiation \cite{Rieth2013,Abernethy2016,Dubinko2017}. Therefore, an understanding of material properties on atomistic level, in particular the mechanisms of embrittlement, is a challenging task which can be achieved by combined use of proper theoretical and experimental methods.  Large-scale atomistic simulations, based on molecular dynamics (MD), are widely used to address this problem, but their reliability depends on the choice of interatomic potentials.
Note that more than 30 different interatomic potentials are available for tungsten nowadays \cite{Bonny2014}.

The accuracy of the interatomic potentials represents often a bottleneck of the MD simulations, therefore their validation becomes crucial.  Structural, thermodynamic and vibrational  properties of a material are conventionally used for this purpose \cite{Gunsteren1990,Choudhary2017}. 
Another source of useful structural and dynamic information 
is the extended X-ray absorption fine structure (EXAFS), 
which includes also contributions from high-order atomic distribution functions, giving origin to the so-called multiple-scattering events \cite{Rehr2000}. 
The first uses of the MD simulations for the interpretation of EXAFS spectra  date back to nineties of the last century \cite{DAngelo1994,DAngelo1996,Palmer1996direct,Kuzmin1997}.  More recently the approach was widely applied to disordered \cite{DAngelo2002,DiCicco2002,Merkling2003,Okamoto2004,Ferlat2005,DAngelo2013,Merat2013,Spezia2017}, nanosized \cite{Anspoks2010nio,Roscioni2011,anspoks2011interpretation,anspoks2012atomic,Price2012,Yancey2013,Chill2015,Timoshenko2017jcp} and crystalline \cite{Anspoks2010nio,kuzmin2009quantum,Kuzmin2011,Timoshenko2011ge,Timoshenko2014zno,Jonane2016fef3,Kalinko2016awo4,Jonane2016} materials. The possibility to use 
EXAFS spectra for the validation of interatomic potentials was demonstrated in \cite{kuzmin2009quantum,Kuzmin2016zpc,Bocharov2017}.  

A complementary atomistic simulation approach to EXAFS spectrum analysis is based 
on  reverse Monte Carlo (RMC) method \cite{Gurman1990,DiCicco2005,Krayzman2008,Krayzman2010,DiCicco2014,bridges2014local,Kompch2015,Gereben2015,Pethes2016,Timoshenko2017nano,Triana2017}.
It was realised in a number of computer codes as RMCprofile
\cite{tucker2007rmcprofile}, SpecSwap-RMC \cite{Leetmaa2010}, RMC++ \cite{gereben2012rmc_pot} and EvAX \cite{timoshenko2014exafs}.
The method was successfully used by us in the analysis of several materials
as perovskites \cite{timoshenko2014exafs,Timoshenko2017cu3n,Jonane2016fef3},
tungstates \cite{timoshenko2015local,timoshenko2016cowo4,Kalinko2016awo4},
ZnO \cite{Timoshenko2014zno} and Y$_2$O$_3$ \cite{Jonane2016}.
While the MD-EXAFS approach deals with a time-dependent 3D model of a material
and allows one to evaluate the configuration-averaged EXAFS spectrum  from a set of atomic coordinates accumulated during the MD run,
the RMC method solves an inverse problem thus reconstructing static atomic configuration from the experimental EXAFS data \cite{Gurman1990,timoshenko2014exafs}.

Here we demonstrate the use of both approaches on the example of the W L$_3$-edge EXAFS spectrum analysis for bcc tungsten.

\section{Experimental}
\label{exper}

Good quality W L$_3$-edge X-ray absorption spectrum of tungsten metallic foil (99.95\%, Goodfellow) was recorded at $T=300$~K  
in transmission mode at the ELETTRA XAFS bending magnet beamline \cite{ELETTRA}. 
The storage ring operated in the top-up multibunch mode at the energy $E=2.4$~GeV and current $I=160$~mA. The synchrotron radiation was monochromatized using a Si(111) double-crystal monochromator, and its intensity before and after the
sample was measured by ionization chambers filled with a mixture of Ar, He and N$_2$  gases. 

The experimental W L$_3$-edge EXAFS spectrum  was extracted using the conventional procedure \cite{Aksenov2006,EDA} and is shown together with its 
Fourier transform (FT) in Fig.~\ref{fig1}. Note that the peaks up to 10~\AA,  due to the nearest 14 coordination shells around 
the absorbing tungsten atom, are clearly visible at $T=300$~K in the FT.

\section{Molecular dynamics simulations}
\label{mdexafs}

Classical MD simulations  were performed in the canonical (NVT) ensemble with periodic boundary conditions  by the GULP4.3 code \cite{gale2003general}.
The simulation box with bcc tungsten crystal structure was a 7$a_0$$\times$7$a_0$$\times$7$a_0$ supercell containing 686 atoms ($a_0 =3.165$~\AA\  \cite{Parrish1960,Dutta1963}).
The Newton's equations of motion were integrated with the Verlet leapfrog algorithm \cite{Hockney1970},
using a time step of 0.5~fs. The Nos\'{e}-Hoover thermostat \cite{Hoover1985} was used to keep the average temperature around $T=300$~K during the simulations.
After equilibration during 20~ps, a set of 4000 static atomic configurations was collected for the next 20~ps.
The MD simulations were performed for two force-field models: the Finnis-Sinclair embedded atom model (EAM) potential \cite{Finnis1984} and the second nearest-neighbor modified embedded atom method (2NN MEAM) potential \cite{Lee2001}.

Sets of static atomic configurations obtained in the MD simulations were used to calculate the
configuration-averaged W L$_3$-edge EXAFS $\chi(k)$ ($k$ is the photoelectron wavenumber) within the multiple-scattering (MS) approach \cite{Kuzmin2016zpc,kuzmin2009quantum,Kuzmin2014} using ab initio self-consistent real-space MS FEFF8.50L code \cite{ankudinov1998,Rehr2000}.
The scattering potential and partial phase shifts were calculated within the muffin-tin (MT)
approximation \cite{ankudinov1998} only once for the cluster with the radius of 10~\AA, centered at the absorbing tungsten atom  and constructed based on the average atomic configuration,  corresponding to bcc tungsten structure. Small variations of the cluster potential due to  thermal vibrations during the MD simulations were neglected. The MS contributions were accounted up to the 6th order to guarantee the convergence of the total EXAFS in the $k$-range of 3--18~\AA$^{-1}$. The photoelectron inelastic losses were accounted within the one-plasmon approximation using the complex exchange-correlation Hedin-Lundqvist potential \cite{Hedin1971}. 
The amplitude reduction factor $S_0^2$ is included in the scattering amplitude \cite{Rehr2000}, calculated by the FEFF code, and no additional correction of the EXAFS amplitude was performed.

\begin{figure}[t]
	\centering
	\includegraphics[width=0.6\textwidth]{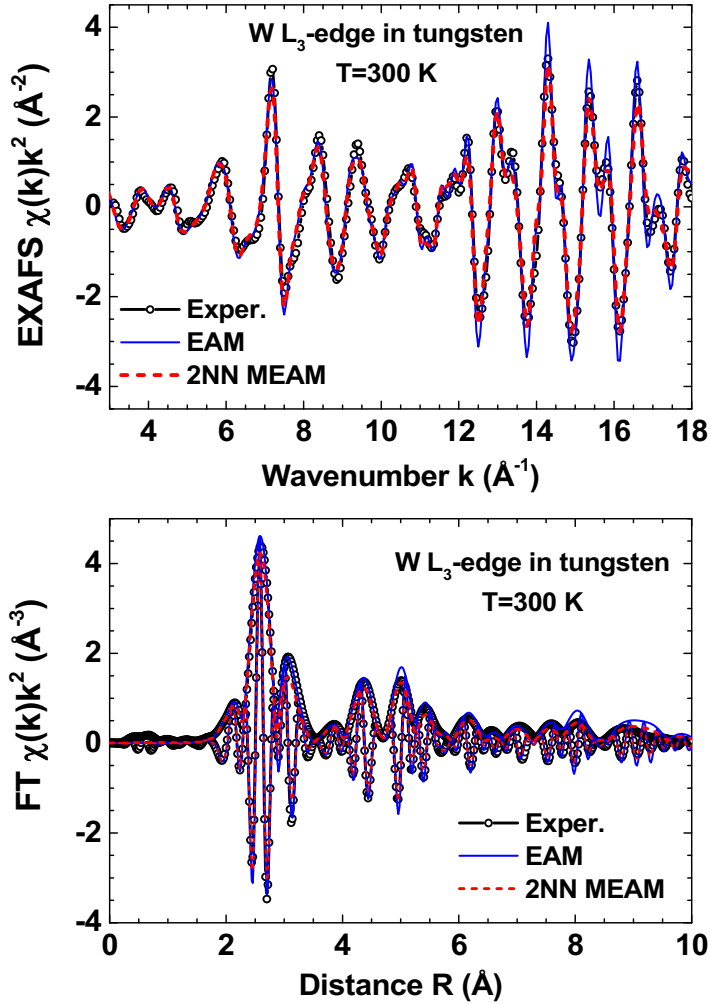}
	\caption{Comparison of the experimental  and calculated W L$_3$-edge MD-EXAFS 
		$\chi(k)k^2$ spectra and their Fourier transforms (FTs) (modulus and imaginary parts are shown) in bcc tungsten at $T=300$~K.  }
	\label{fig1}
\end{figure}

\begin{figure}[t]
	\centering
	\includegraphics[width=0.6\textwidth]{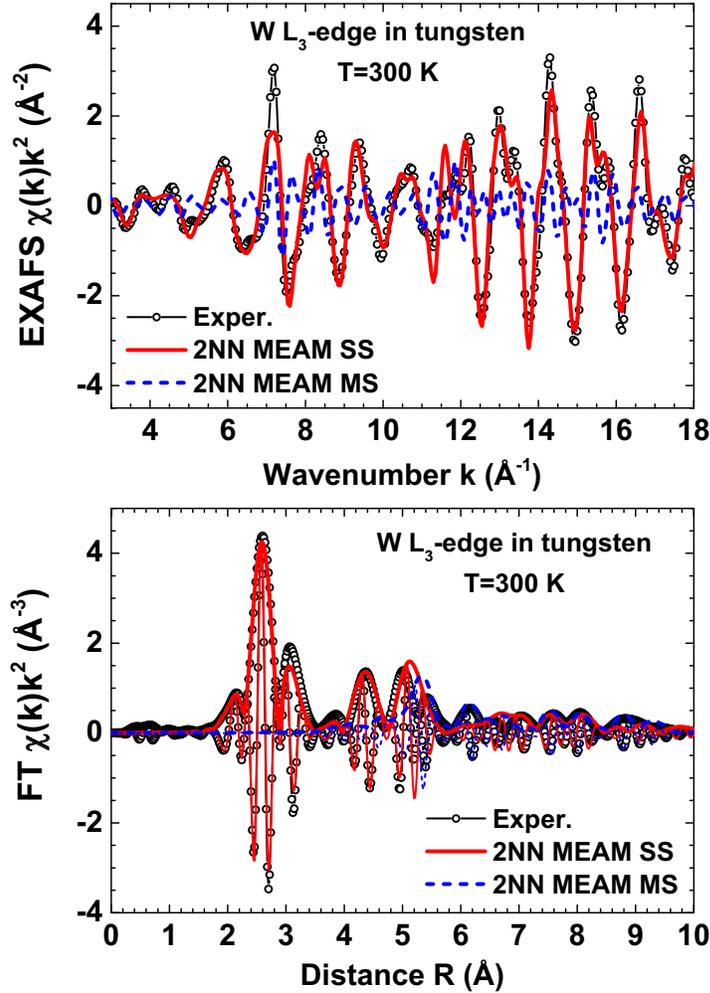}
	\caption{ A sum of the single-scattering (SS) and multiple-scattering (MS) contributions 
		to the W L$_3$-edge MD-EXAFS $\chi(k)k^2$ spectra (2NN MEAM potential) and their Fourier transforms (FTs) 
		(modulus and imaginary parts are shown) for bcc tungsten at $T=300$~K. Open circles -- the experimental data.}
	\label{fig2}
\end{figure}

\begin{figure}[t]
	\centering
	\includegraphics[width=0.6\textwidth]{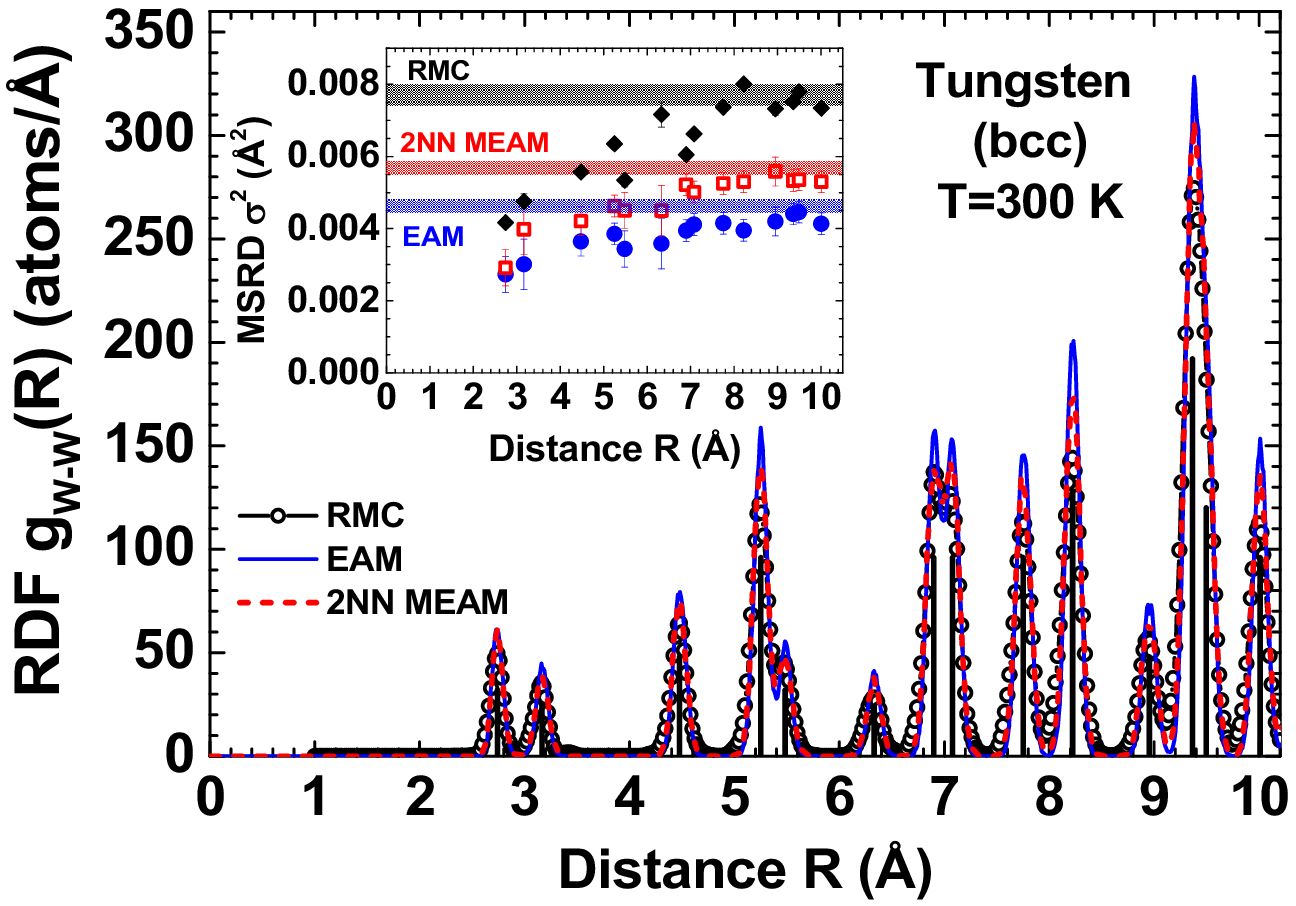}
	\caption{ Radial distribution functions (RDFs) $G_{\rm W-W}(R)$ 
		calculated from the results of the MD and RMC/EA simulations for bcc tungsten at $T=300$~K.
		Vertical lines show crystallographic data.	
		Inset: Dependence of the MSRD $\sigma^2_{\rm W-W}(R)$ on distance.
		Horizontal lines correspond to a sum of two MSDs of  tungsten. }
	\label{fig3}
\end{figure}

The configuration-averaged W L$_3$-edge EXAFS spectra and their Fourier transforms (FTs) are compared with the experimental data in $k$ and $R$ spaces in Fig.~\ref{fig1}. The single-scattering (SS) and MS contributions to the total EXAFS spectrum and
their FTs are shown in Fig.~\ref{fig2}. The total and partial radial distribution functions (RDFs) $G(R)$ obtained from the MD simulations are reported in Fig.~\ref{fig3}. The mean-square relative displacements (MSRD) $\sigma^2_{\rm W-W}(R)$ and mean-square displacements (MSDs) of tungsten atoms were also evaluated (see the inset in Fig.~\ref{fig3}).

%
%

%
%

\section{Reverse Monte Carlo simulations}
\label{rmc}

RMC method based on the evolutionary algorithm  (EA), implemented in the EvAX code \cite{timoshenko2014exafs}, was used to obtain structural model of bcc tungsten consistent with the 
experimental W L$_3$-edge EXAFS spectrum.
The simulation box was a 5$a_0$$\times$5$a_0$$\times$5$a_0$ supercell (250 atoms) with periodic boundary conditions. Starting atomic configuration was constructed according to the diffraction data \cite{Parrish1960,Dutta1963}. RMC/EA calculations were simultaneously performed for 32 atomic configurations.  At each iteration new atomic configuration was generated by randomly displacing all atoms in the box with the maximal allowed shift of 0.4~\AA\  to get best possible agreement between the Morlet wavelet transforms (WTs) of the experimental and theoretically calculated W L$_3$-edge EXAFS spectra \cite{timoshenko2009wavelet,timoshenko2012reverse}. 
No significant improvement in the residual was observed after 4000 iterations.
As in MD-EXAFS simulations, the configuration-averaged EXAFS spectra were calculated by ab initio real-space FEFF8.50L code \cite{ankudinov1998,Rehr2000} including MS effects up to 6th order. Calculations were performed in the $k$-space range from 3 to 18~\AA$^{-1}$ and in the $R$-space range from 1 to 8~\AA. 
The amplitude reduction factor $S_0^2$ is included in the scattering amplitude \cite{Rehr2000}, calculated by the FEFF code, and no additional correction of the EXAFS amplitude was performed.

\begin{figure*}[t]
	\centering
	\includegraphics[width=0.95\textwidth]{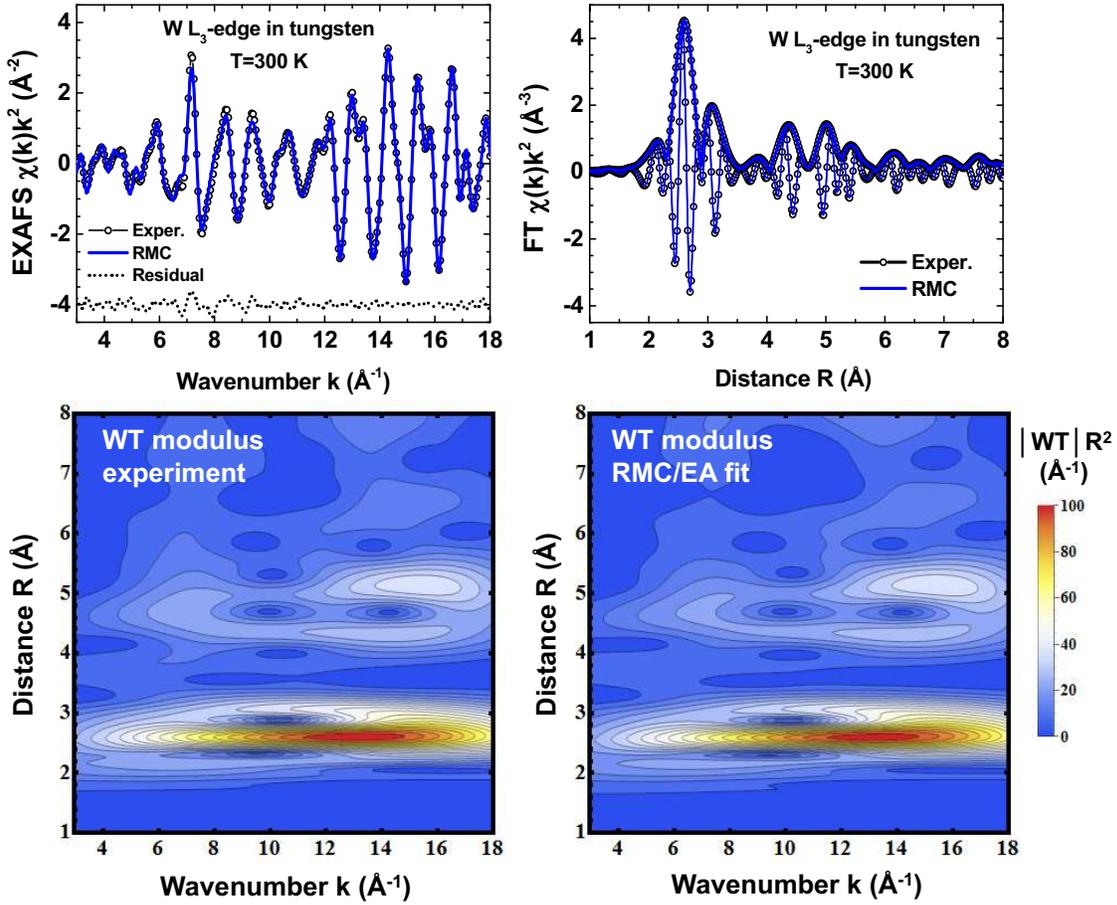}
	\caption{ Comparison of the experimental  and RMC/EA calculated  
		W L$_3$-edge EXAFS spectra $\chi(k)k^2$ and their Fourier transforms (FTs)
		(upper panels)  as well as the moduli of their Morlet wavelet transforms (WTs) 
		(lower panels)  in bcc tungsten at $T=300$~K. The residual between the experimental and calculated EXAFS  spectra is shown by dotted line. Both modulus and imaginary parts are shown for FTs.}
	\label{fig4}
\end{figure*}

The result of the RMC/EA calculations is shown in $k$, $R$ and WT spaces in Fig.~\ref{fig4}. The average RDF function, which was calculated from atomic coordinates obtained in two RMC/EA simulations with different sets of pseudo-random numbers \cite{timoshenko2014exafs}, is compared with the MD results in Fig.~\ref{fig3}.

\section{Results and discussion}
\label{results}

\begin{figure*}[t]
	\centering
	\includegraphics[width=0.6\textwidth]{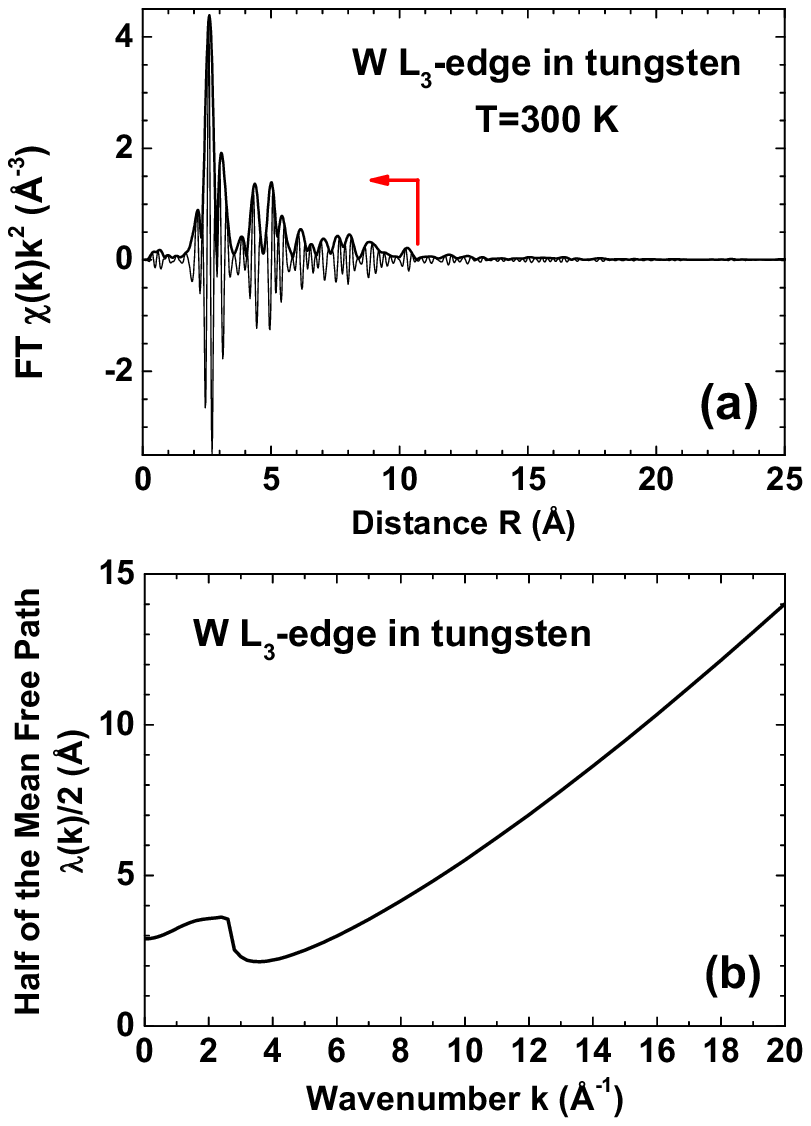}
	\caption{ (a) Fourier transform of the experimental W L$_3$-edge EXAFS spectrum $\chi(k)k^2$ 
		in bcc tungsten at $T=300$~K. The region of structural peaks is indicated by arrow. 
		(b) Calculated photoelectron mean free path (MFP) $\lambda(k)$ in bcc tungsten.}
	\label{fig5}
\end{figure*}

It is commonly believed that EXAFS spectroscopy is a local structural method. 
However, one can ask a question: how local is EXAFS spectroscopy? 
The region of a structure around the absorber contributing into EXAFS is determined by the photoelectron mean free path (MFP) and core hole lifetime. Additionally, the EXAFS is dumped by structural and thermal disorder, which are material dependent. In the case of crystalline materials with well-ordered high symmetry structure, enough strong bonds and consisting of chemical elements with scattering amplitude maxima located within the measured EXAFS $k$-range, the contributions from distant shells (up to 8-10~\AA) can be observed in the high quality experimental EXAFS spectra. For example, the structural peaks were found up to about 9.5~\AA\ in the Ni K-edge EXAFS of cubic rock-salt NiO \cite{Kuzmin2014}. Similar situation occurs in metallic bcc tungsten, which has strong scattering amplitude at large $k$-values (Fig.~\ref{fig5}). For tungsten, the calculated MFP $\lambda(k)/2 \simeq 12$~\AA\  at $k_{\rm max}=18$~\AA$^{-1}$. This result is consistent with the fact that structural peaks are detectable in FT of the experimental W L$_3$-edge EXAFS up to about 10.5~\AA. The structural origin of these peaks is
supported by an agreement with the model MD-EXAFS spectra in Fig.~\ref{fig1}.

Classical MD NVT simulations performed with EAM \cite{Finnis1984} and 2NN MEAM \cite{Lee2001} potentials result in a set of atomic configurations, which
allow generating configuration-averaged EXAFS spectra in agreement with
the experimental data (Fig.~\ref{fig1}). Position and amplitude of the most peaks up to 10~\AA\ are well reproduced suggesting that both potential models describe well
atomic structure  and thermal disorder in bcc tungsten at $T=300$~K.
However, the detailed comparison of the  W L$_3$-edge EXAFS spectra in $k$-space indicates that
the residual between the experimental and MD-EXAFS spectra is about twice smaller in the case of 
the simulation based on the 2NN MEAM potential. Finally, best agreement with the experimental EXAFS data was obtained by the RMC/EA approach (Fig.~\ref{fig4}).

Comparison of partial SS (due to pair correlations) and MS (due to many atom correlations) contributions to the total W L$_3$-edge EXAFS, obtained using the MD-EXAFS simulation with the 2NN MEAM potential \cite{Lee2001}, is shown in $k$ and $R$ space in Fig.~\ref{fig2}. 
Significant MS contribution is present in the whole $k$-range and above 4~\AA\ in $R$-space. This means that EXAFS signal from only first two coordination shells of tungsten (peaks from 1.5 to 3.5~\AA) can be  accurately analysed within the SS approximation. For outer shells the MS signals produce comparable or even dominating contribution to the total EXAFS spectrum, and, thus, should be taken into account.

RDFs $G_{\rm W-W}(R)$ obtained for both potential models and from the RMC/EA fit of the EXAFS data are shown  in Fig.~\ref{fig3}. The widths of the peaks in RDFs determine
the magnitude of disorder described by the MSRD $\sigma^2_{\rm W-W}(R)$ parameters (see the inset).   
The smallest MSRD values were found for the EAM potential, while slightly larger 
MSRD values were obtained for the 2NN MEAM potential. The strongest broadening 
of the peaks was observed for the RMC/EA result, however, it is known that the 
RMC method gives a solution with maximal disorder among all possible structure 
models \cite{Timoshenko2016cu3n}. 
Anyway in all three cases the MSRD $\sigma^2_{\rm W-W}(R)$  approaches the sum of two
MSD values, shown by horizontal lines in Fig.~\ref{fig3}, 
at large distances ($R \gtrsim 8$~\AA), i.e. for distant coordination shells.
The MSRD and MSD values for the $i$-$j$ atom pair are
related as ${\rm MSRD}_{ij} = {\rm MSD}_i + {\rm MSD}_j - 2  \varphi \sqrt{{\rm MSD}_i} \sqrt{{\rm MSD}_j}$,
where $\varphi$ is a dimensionless correlation parameter \cite{Booth1995}.
Therefore, the behaviour of the W--W MSRD $\sigma^2_{\rm W-W}(R)$  at large distances in Fig.~\ref{fig3}
reflects disappearance of correlations $\varphi$\ in atomic motion
of distant tungsten atoms \cite{Jonane2016,Sapelkin2002,Jeong2003}.
This means that EXAFS data can be used to obtain both MSRD and MSD values,
if information from distant shells is available and can be extracted.  

Finally, one can compare the values of the MSDs for tungsten obtained by MD and RMC simulations (see inset in Fig.~\ref{fig3}) with those 
determined from x-ray diffraction experiments \cite{Houska1964,Paakkari1974} and calculated from lattice dynamics \cite{Fine1939,Dobrzynski1972}. Our simulations predict MSD(EAM)$=$0.0023~\AA$^2$, MSD(MEAM)$=$0.0029~\AA$^2$ 
and MSD(RMC)$=$0.0039~\AA$^2$ in reasonable agreement compared to the experimental MSD$=$0.0061~\AA$^2$ in \cite{Houska1964} and 0.0022~\AA$^2$ in \cite{Paakkari1974} and the calculated MSD=0.0023~\AA$^2$ in \cite{Fine1939} and 0.0018~\AA$^2$ in \cite{Dobrzynski1972}.

\section{Conclusions}
\label{conc}

In this study we performed atomistic simulations of the experimental W L$_3$-edge extended X-ray absorption fine structure (EXAFS) spectrum of bcc tungsten at $T=300$~K using two complementary approaches -- molecular dynamics (MD) and reverse Monte Carlo (RMC) methods. High quality of the experimental W L$_3$-edge EXAFS spectrum and the use of two advanced approaches allowed us to extend analysis far beyond the first coordination shell. 
Contributions from outer coordination  shells within the range defined by the photoelectron mean free path (Fig.~\ref{fig5})  are well visible in the EXAFS of bcc tungsten due to its well-ordered high symmetry structure and strong backscattering amplitude.  

Classical MD simulations were conducted in the canonical (NVT) ensemble for two force-field  models -- EAM \cite{Finnis1984} and 2NN MEAM \cite{Lee2001}, and the configuration-averaged EXAFS spectra were calculated within the MD-EXAFS approach \cite{kuzmin2009quantum,Kuzmin2016zpc} based on ab initio multiple-scattering formalism  \cite{Rehr2000,ankudinov1998}.
The obtained results suggest that both force-field  models allow one to reproduce well the experimental W L$_3$-edge EXAFS spectrum of bcc tunsgten, however the simulation using the 2NN MEAM potential results in about twice smaller residual. We have shown that multiple-scattering contributions become important starting from the third coordination shell and should be accounted in the analysis. RMC analysis gives best agreement with the experimental EXAFS data and predicts slightly larger MSRD values for all coordination shells of tungsten than both MD simulations. 

The possibility to analyse the W L$_3$-edge EXAFS of bcc tungsten from distant coordination shells up to $\sim$10~\AA\ was demonstrated and allowed us to
extract the MSRD $\sigma^2_{\rm W-W}(R)$ dependence (Fig.~\ref{fig3}). We found that
the correlation in atomic motion in bcc tungsten becomes negligible above 8~\AA, so that the MSRD values approach a sum of two MSD factors. The obtained values of MSD 
are in reasonable agreement with limited number of available data \cite{Houska1964,Paakkari1974,Fine1939,Dobrzynski1972}.
This fact indicates that the analysis of outer shell contributions allows one to estimate the MSD factors directly from EXAFS data.

\ack
The authors gratefully acknowledge the assistance of the ELETTRA XAFS beamline staff members during the EXAFS experiment No 20150303.
This work has been carried out within the framework of the EUROfusion Consortium and has received funding from the Euratom research and 
training programme 2014-2018 under grant agreement No 633053. The views and opinions expressed herein do not necessarily reflect those of the European Commission.

\section*{References}

\providecommand{\newblock}{}

\end{document}